\documentclass[reprint,
 amsmath,amssymb,
 aps,
]{revtex4-2}

\usepackage{graphicx}
\usepackage{dcolumn}
\usepackage{bm}
\usepackage{amsmath,amssymb}
\usepackage{color}
\usepackage{romannum}
\usepackage{hyperref}

\begin{document}

\preprint{APS/123-QED}

\title{Dynamics of Poro-viscoelastic Wetting with Large Swelling}


\author{B. X. Zheng}
\affiliation{Mechanics Division, Department of Mathematics, University of Oslo, 0316 Oslo, Norway.}%

\author{T. S. Chan}
\email{taksc@uio.no}
\affiliation{Mechanics Division, Department of Mathematics, University of Oslo, 0316 Oslo, Norway.}

\author{E. H. van Brummelen}
\email{e.h.v.brummelen@tue.nl}
\affiliation{Multiscale Engineering Fluid Dynamics Group, Department of Mechanical Engineering, Eindhoven University of Technology,
P.O. Box 513, 5600 MB Eindhoven, The Netherlands}
 
\author{J. H. Snoeijer}%
\email{j.h.snoeijer@utwente.nl}
\affiliation{Physics of Fluids Group, Faculty of Science and Technology, University of Twente, 7500 AE Enschede, The Netherlands
}%

\begin{abstract}
The deposition of droplets onto a swollen polymer network induces the formation of a wetting ridge at the contact line. Current models typically consider either viscoelastic effects or poroelastic effects, while polymeric gels often exhibit both properties. In this study, we investigate the growth of the wetting ridge using a comprehensive large-deformation theory that integrates both dissipative mechanisms - viscoelasticity and poroelasticity. In the purely poroelastic case, 
following an initial instantaneous incompressible deformation, the growth dynamics exhibit scale-free behavior, independent of the elastocapillary length or system size. A boundary layer of solvent imbibition  between the solid surface (in contact with the reservoir) and the region of minimal chemical potential is created.  At later times, the ridge equilibrates on the diffusion timescale given by the elastocapillary length.
When viscoelastic properties are incorporated, our findings show that, during the early stages (prior to the viscoelastic relaxation timescale), viscoelastic effects dominate the growth dynamics of the ridge and solvent transport is significantly suppressed. Beyond the relaxation time, the late-time dynamics closely resemble those of the purely poroelastic case. These findings are discussed in light of recent experiments, showing how our approach offers a new interpretation framework for wetting of polymer networks of increasing complexity.
\end{abstract}

\maketitle


\section{\label{sec:level1}Introduction}

The wetting of reticulated polymer networks, such as gels, involves phenomena with far-reaching implications across diverse fields. It plays a significant role in the design of biomimetic materials \cite{sidorenko2008controlled, tam2011factors} and self-healing polymeric materials \cite{zheng2022self, wool1981theory}. Understanding the influence of capillary forces on solvent transport and hydrogel deformation is crucial for optimizing applications such as drug delivery, tissue engineering, and soft adhesion systems \cite{Ullah2015,Antonio2009,TianyiZhao2018}. Additionally, wetting is essential for understanding biophysical processes, including bleb formation in cells \cite{charras2005non}, and skin overhydration \cite{cutting2002maceration}.  Despite its broad range of applications, contemporary understanding of the physical behavior of gels in wetting scenarios is still incomplete.

One fundamental wetting scenario involves placing a droplet on a solid surface, offering a model system for both theoretical and experimental studies to address key questions in soft wetting \cite{Carre1996,Pericet-Camara2008,Jerison2011,Yu2012,Style2012,Kajiya2013,park2014visualization,Style2017,andreotti2020statics, khattak2022direct}. When a droplet contacts a soft substrate, capillary forces pull the solid at the contact line to form a wetting ridge  \cite{Lester1961,Rusanov1975,Carre1996,Pericet-Camara2008,Yu2009,Pericet-Camara2009,Jerison2011,Yu2012,Style2012,Limat2012,style2012universal,Kajiya2013, park2014visualization,lubbers2014drops,Dervaux2015, karpitschka2015droplets,Style2017,andreotti2020statics,pandey2020singular,Chan2022,Zheng2023,Brummelen:2021wt,BXue2025}. This situation is shown schematically in Fig.~\ref{fig:epsart}(a,b).  While most studies focused on the static, equilibrium properties of the wetting ridge, their dynamical behavior can be intricate.  The growth dynamics of the ridge essentially involves two possible dissipative mechanisms. First, the rearrangement of the polymer chains within the network gives rise to time-dependent elastic properties, with the process characterized by intrinsic viscoelastic relaxation timescales \cite{long1996static, Carre1996, karpitschka2015droplets, zhao2018geometrical, andreotti2020statics}. Second, polymeric gels can absorb solvent and swell \cite{ZhaoMenghua2018,xu2020viscoelastic}. The transport of solvent follows a diffusive process characterized by poroelastic timescales. These two dissipation mechanisms -- viscoelasticity and poroelasticity -- can occur at the same time and are intricately coupled. 

Various studies have aimed at unraveling the dynamics of poroelastic gels. Nevertheless, the interpretation of experimental results remains ambiguous. Using a linear model of poroelasticity, Zhao et al. \cite{ZhaoMenghua2018} demonstrated that the ridge exhibits logarithmic growth over time. Instead of probing the growth dynamics, Xu et al. \cite{xu2020viscoelastic} investigated the relaxation of an equilibrium ridge upon the removal of the contacting droplet. Their experimental results suggest that the early-time dynamics are dominated by viscoelastic effects, while poroelasticity plays a main role in the late-time behavior. This conclusion contrasts a study based on Atomic Force Microscopy by Li et al. \cite{li2023crossover}, which suggests that the stress relaxation of hydrogels undergoes a crossover from short-time poroelastic relaxation to long-time viscoelastic relaxation. Additionally, an alternative investigation of soft-gel dynamics, conducted by Berman et al. \cite{berman2019singular}, focused on the process of adhesive detachment. The authors showed that the early-time relaxation rate is primarily governed by porous flow driven by solid surface stresses. Returning to the context of droplet wetting, recent experimental studies have revealed distinct behaviors of solvent distribution within the ridge region of a swollen gel. Cai et al. \cite{cai2021fluid} demonstrated that soft PDMS gels phase separate from their solvent in the contact line region. Similar phase separation has also been observed in adhesive contact between a solid bead and a silicone gel substrate \cite{jensen2015wetting}. 

The incongruent findings highlight the complex interplay between viscoelastic and poroelastic behaviors in soft gels. Yet,  a comprehensive theory that integrates both dissipative mechanisms in wetting scenarios is still lacking. Furthermore, models used to understand ridge dynamics often rely on the assumption of small deformation \cite{Style2012,ZhaoMenghua2018}, although deformations are large and stresses tend to diverge upon approaching the contact line \cite{pandey2020singular}. The necessity of using a large deformation theory was further evidenced by a recent theoretical work by Flapper et al. \cite{flapper2023reversal}, which reveals a significantly different equilibrium swelling within the wetting ridge compared to results from a linear theory \cite{ZhaoMenghua2018}. 

In this study, we develop a theoretical framework to describe the process of ridge formation and capillary-induced solvent transport in polymeric gels with large deformation and significant swelling. We focus on the growth of the elastocapillary ridge on a substrate, after the application of a point force at $t=0$; see Fig.~\ref{fig:epsart}(b). In particular, we address the key question of how this growth is influenced by the competition of viscoelastic and poroelastic dissipation mechanisms. The subsequent sections will be organized as follows: Section II introduces the formulations for the capillary force-induced gel deformation and solvent transport system. Section III analyzes the results for the case in which the solid is a purely poroelastic material (i.e. without viscoelaticity). Section IV presents the analysis of results that incorporate viscoelastic effects within the system. Finally, conclusions will be drawn in Section V.

\section{\label{sec:level1}Poro-visco-elasto-capillary formulation}

\begin{figure*}[ht]
\includegraphics[width=1\textwidth]{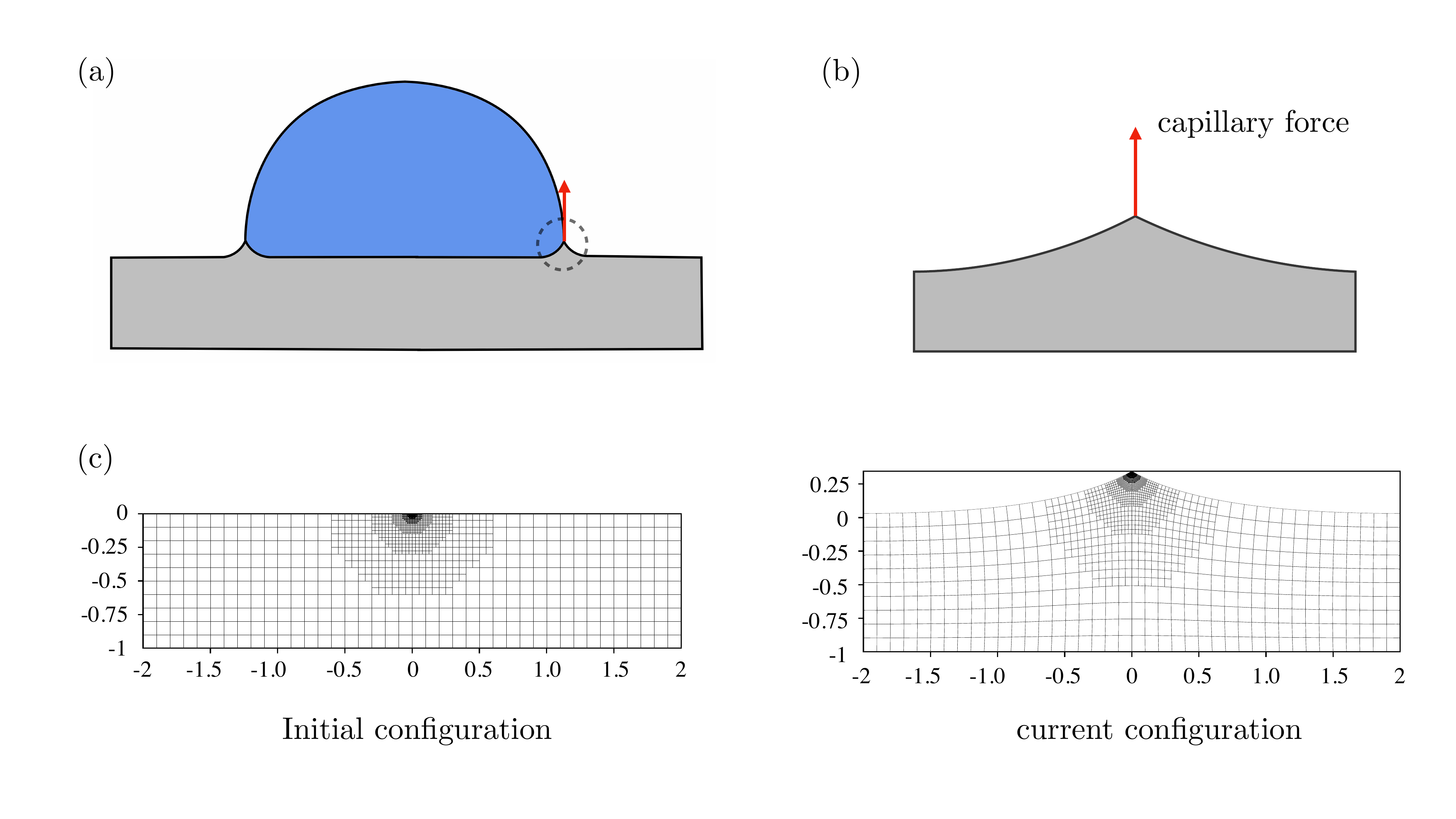}
\caption{\label{fig:epsart} Schematic representation of the physical model. (a) Wetting behavior on a polymeric gel; (b) A simplified model illustrating the capillary force acting on the polymeric gel; (c) Initial configuration demonstrating the finite element mesh for a preswollen polymeric gel; (d) The mesh configuration of the deformed polymeric gel under the influence of capillary force.}
\end{figure*}

In this section, we outline the modeling approach used in this paper, which incorporates poroelasticity, viscoelasticity, and capillarity. The substrate is modeled using the finite deformation poroelastic model by Hong {\rm et al} \cite{hong2008theory}, complemented with a capillary description as in \cite{pandey2020singular,henkel2022soft}. Both descriptions are based on a free-energy formulation, which needs to be complemented with dissipative transport equations. For solvent transport, we adapt a standard diffusive transport model \cite{hong2008theory}, while we introduce viscoelasticity of the polymer network as an additional dissipative mechanism. This enables us to address the question of whether/when the dynamics is governed by solvent transport or by viscoelastic relaxation. As is common in the study of contact line dynamics, we adopt a two-dimensional description where the action of the contact line is modelled by a line force (Fig.~\ref{fig:epsart}(b)). This reduction reflects that under typical experimental conditions the droplet size is generally much larger than the deformation scale near the contact line (Fig.~\ref{fig:epsart}(a)). Consequently, this allows the use of a plane strain description to formulate the problem.

\subsection{\label{sec:level2}Poro-elastocapillary substrate}

The deformation of the polymer matrix is characterized by a mapping from a reference configuration prior to deformation (i.e. a dry gel without any solvent) to the current state. By standard notation, this mapping is expressed as 
\begin{equation}
\label{eq:mapping}
\mathbf{x}=\boldsymbol{\psi}(\mathbf{X}, t).
\end{equation}
Here, $\mathbf{X}$ represents the position of a material point in the reference domain, which is mapped to its current position $\mathbf{x}$ by the deformation map $\boldsymbol{\psi}$. We assume that the geometry is invariant along the contact line, allowing us to treat the problem as two-dimensional (plane strain elasticity). The polymer matrix can be swollen by a solvent, so the mapping $\boldsymbol{\psi}$ does not preserve volume. Defining the deformation gradient tensor as $\mathbf F =\partial \mathbf{x} / \partial \mathbf{X}$, the relative change in volume is given by 
\begin{equation}
\label{eq:swelling ratio}
J=\mathrm{det}(\mathbf F).
\end{equation}
Using the dry network without any solvent as the reference configuration, $J$ naturally defines the ``swelling ratio". Next, it is assumed that both the polymer molecules that make up the matrix and the solvent molecules are incompressible. This assumption implies that the volumetric change of the network directly determines the local number density of the solvent molecules. Defining $C(\mathbf X,t)$ as the number of solvent molecules per unit volume in the reference configuration, the molecular incompressibility is ensured by the constraint~\cite{hong2008theory}
\begin{equation}
\label{eq:incompressibility}
J = 1+v C,
\end{equation}
where $v$ is the volume per solvent molecule. The system is thus described by two fields $\boldsymbol{\psi}(\mathbf X,t)$ and $C(\mathbf X,t)$, which are constrained by (\ref{eq:incompressibility}).  

The free energy of the polymer matrix is described using a hyperelastic formulation based on a strain-energy density $W_{\mathrm{el}}(\mathbf{F})$, per unit volume in the reference configuration. In the present study, we adopt a common neo-Hookean model based on Gaussian chain elasticity, which for plain-strain conditions gives the energy density
\begin{equation}\label{eq:neohookean}
W_{\mathrm{el}}(\mathbf{F})=\frac{1}{2} G[\mathbf{F}: \mathbf{F}-2-2 \log (\operatorname{det}(\mathbf{F}))],
\end{equation}
where $G$ is the shear modulus of the network. The network elasticity is of entropic origin, which can be appreciated from the connection $G=NkT$, where $N$ represents the number of chains per unit volume \cite{boyce2000constitutive, marckmann2006comparison, hong2008theory,kim2022polyacrylamide}, and $kT$ is the thermal energy. The free energy for the mixing of the solvent and polymer can be described using the Flory--Huggins theory. The corresponding energy density with respect to volume measure in the reference configuration takes the form~\cite{hong2008theory}
\begin{equation}\label{eq:floryhuggins}
W_{\textsc{fh}}(C)=-\frac{k T}{v}\left[v C \log \left(1+\frac{1}{v C}\right)+\frac{\chi}{1+v C}\right].
\end{equation}
The first term represents the entropy of mixing, while the second term is the enthalpy of mixing, the strength of which is set by the interaction parameter $\chi$. The free-energy density for the swollen poroelastic network is comprised of the sum $W_{\mathrm{el}}(\mathbf{F})+W_{\textsc{fh}}(C)$.


Next, we address the capillary contributions. These contributions come in two forms. First, we endow the surface of the substrate with a surface energy $\gamma_s$. Adopting the convention that is common in capillarity, this surface free energy is measured per unit area in the current configuration. In principle this energy can be a function of the degree of swelling at the interface, which would give rise to the Shuttleworth effect \cite{pandey2020singular,henkel2022soft}; for simplicity it is assumed that $\gamma_s$ is independent of  deformation. 
The second capillary effect is due to the liquid-vapor surface tension 
$\gamma_{\textsc{lv}}$. This can be represented by a localized force \cite{andreotti2020statics}, pulling along the liquid-vapor interface, and leads to the formation of the elasto-capillary ridge as indicated in Fig.~\ref{fig:epsart}(b). Accordingly, we introduce a perfectly localized external traction, characterized by the associated work functional 
$\mathcal{W}(\boldsymbol{\psi})=\gamma_{\textsc{lv}} \mathbf{t}_{\textsc{lv}}  \cdot \boldsymbol{\psi}(\mathbf{X}_{\mathrm{cl}})$,
where $ \mathbf{X}_{\mathrm{cl}}$ denotes the position of the contact line on the solid surface in the reference configuration, and $\mathbf 
    t_{\textsc{lv}}$ is the pulling direction. 

In summary, the functionals describing the free energy 
of the poro-elastocapillary substrate, and the work performed by the liquid-vapour surface tension, per unit length along the contact line, are given by
\begin{subequations}
\label{eq:ewfuncs}
\begin{align}
\mathcal{E}[\boldsymbol{\psi},C,\Pi]
&=
\int{}\mathrm{d}^2X\,\big(W_{\mathrm{el}}+W_{\textsc{fh}}
+
\Pi(1+vC-J)\big)
\notag
\\
&\phantom{=}
+\int{}\mathrm{d}s\,\gamma_s
\label{eq:free_energy}
\\
\mathcal{W}[\boldsymbol{\psi}]&=\gamma_{\textsc{lv}} \mathbf{t}_{\textsc{lv}}  \cdot \boldsymbol{\psi}(\mathbf{X}_{\mathrm{cl}})
\label{eq:work}
\end{align}
\end{subequations}
respectively, where we incorporated the incompressibility constraint~\eqref{eq:incompressibility} into the free-energy function via the Lagrange multiplier~$\Pi$. The integration measure $\mathrm{d}s$ represents the arc length in the deformed configuration.

In the model that we consider, the poroelastic substrate is contiguous to a liquid reservoir. The absorption of solvent molecules into and deformation of the poroelastic substrate also affect the energy of this liquid reservoir. We assume that during its interaction with the substrate, the liquid reservoir remains at constant pressure, $p_0$, and chemical potential, $\mu_0$. Accordingly, the deformation of the substrate produces additional energy in the amount of $p_0V$ via work on the reservoir, and the absorption of solvent molecules consumes energy in the amount $\mu_0N_c$, where $V$ and $N_c$ denote the volume of the substrate and the number of solvent molecules it contains, respectively. 
Noting that $C=(J-1)/v$ by~\eqref{eq:incompressibility}, the energetic contributions from the interaction of the substrate with the liquid reservoir can be expressed in terms of the deformation and concentration as
\begin{equation}
\label{eq:Ereservoir}
\begin{aligned}
p_0V-\mu_0N_c
&=
p_0\int{}\mathrm{d} ^2X\,J-\mu_0\int{}\mathrm{d}  ^2X\,C    
\\
&=
\tilde{p}_0\int{}\mathrm{d} ^2X\,J+\operatorname{const}
\end{aligned}
\end{equation}
with $\tilde{p}_0=p_0-\mu_0/v$. The identities in~\eqref{eq:Ereservoir} convey that for an incompressible system, an exchange of molecules cannot be distinguished from an exchange of volume. Hence, without loss of generality, we will in the remainder set the chemical potential of the reservoir 
to~$\mu_0=0$. For an incompressible system, the value of $\tilde p_0$ is without any consequence as well, as it only sets the gauge for the stress. By setting $p_0=0$, it is thus understood that we measure the pressure with respect to that of the reservoir. Hence, in the remainder it will suffice to ignore the energetic contributions of the reservoir, under the convention that the gauge of the reservoir is chosen with $\mu_0=0$ and $p_0=0$.



\subsection{\label{sec:citeref}Dynamical equations and dissipation}
We now consider the evolution equations that model the dynamics of wetting ridges in the presence of poroelasticity and viscoelasticity. In modeling the dynamics of poro-viscoelastic materials, it is generally appropriate to neglect inertial effects. Therefore, we consider the momentum balance in quasi-stationary form, and the solvent evolution in time-dependent form.

We present the conservation equation for the solvent in the current configuration. To this end, we denote by $\mathbf{v}(\mathbf{x},t)=\partial_t\boldsymbol{\psi}(\mathbf{X},t)$ the material velocity of the substrate, 
and by $c(\mathbf{x},t)=C(\mathbf{X},t)/J(\mathbf{X},t)$ the solvent concentration in the current configuration, under the deformation in~\eqref{eq:mapping}.
Conservation of the solvent is then described by a conservation equation of the form
\begin{equation}
\label{eq:dcdt}
\partial_tc+\nabla{}\cdot(c\mathbf{v})+\nabla \cdot \mathbf{q}_c=0
\end{equation}
where $\nabla$ represents the gradient operator in the current configuration and $\mathbf{q}_c$ denotes the diffusive flux. To provide a constitutive relation for the diffusive flux, we introduce the chemical potential of the solvent in the reference configuration according to
\begin{equation}
\label{eq:chempot}    
M:=\frac{\delta\mathcal{E}}{\delta{}C}
=
W'_{\textsc{fh}}+\Pi{}v
\end{equation}
where $(\cdot)'$ denotes differentiation. The chemical potential in the current configuration is then given by $\mu(\mathbf{x},t)=M(\mathbf{X},t)$. We define the diffusive flux as
\begin{equation}
\label{eq:qc}
\mathbf{q}_c=-D\nabla\mu    
\end{equation}
with a constant mobility parameter $D>0$. The mobility governs the diffusive transport of solvent, and will in the remainder be referred to as a diffusion coefficient. More sophisticated models of the diffusive process can be realized by replacing~$D$ with a symmetric-positive-definite tensor $\mathbf{D}$, possibly dependent on~$\boldsymbol{\psi}$ and~$C$. 

Neglecting inertial effects in the motion of the substrate, conservation of linear momentum is expressed by the static-equilibrium relation 
\begin{equation}
\label{eq:divsigma}
\nabla\cdot\boldsymbol{\sigma}=0,
\end{equation}
with $\boldsymbol{\sigma}$ the Cauchy stress tensor.  We consider constitutive relations for $\boldsymbol{\sigma}=\boldsymbol{\sigma}_{\text{el}} + \boldsymbol{\sigma}_{\text{v}}$, that comprise an elastic part $\boldsymbol{\sigma}_{\text{el}}$ and (potentially) a viscous part $\boldsymbol{\sigma}_{\text{v}}$. For the considered neo-Hookean hyperelastic material model, the elastic Cauchy stress tensor follows from the free energy~as 
\begin{equation}
\label{eq:elastic stress}
\boldsymbol{\sigma}_{\text{el}}(\mathbf{x},t)=\big[J^{-1}\mathbf{P}\cdot \mathbf{F}^{T}-\Pi\mathbf{I}\big](\mathbf{X},t)
\end{equation}
with $\mathbf{P}:=\mathrm{d}W_{\text{el}}(\mathbf{F})/\mathrm{d}\mathbf{F}$ as the first Piola--Kirchhoff stress tensor corresponding to the elastic part. 
Note that we have integrated the pressure part $\Pi\mathbf{I}$ originating from the constraint (\ref{eq:incompressibility}) into the elastic Cauchy stress. The same Lagrange multiplier $\Pi$ appears in the chemical potential~(\ref{eq:chempot}), and lies at the origin of the poroelastic coupling. For the viscous part, we adopt a Newtonian stress--strain-rate relation.
\begin{equation}
\label{eq:viscoelastic}
\boldsymbol{\sigma}_{\text{v}}(\mathbf{x},t)
=
\eta\big(\nabla \mathbf{v} + (\nabla \mathbf{v})^T\big)
\end{equation}
with $\eta>0$ as the viscosity. The combined elastic and viscous stress tensors provide a Kelvin--Voigt visco-elastic model.

At the interface between the poroelastic substrate and the liquid reservoir,
the traction exerted by the elastic and viscous stresses are balanced by the surface tension. This dynamic interface condition is encoded as
\begin{equation}
\label{eq:dyncon}
\boldsymbol{\sigma}\cdot \mathbf{n}
=\gamma_s\kappa\mathbf{n}
\end{equation}
where $\mathbf{n}$ denotes the exterior unit normal vector and $\kappa$ denotes the curvature of the interface, under the convention that curvature is negative if the osculating circle is located in the substrate. Condition~\eqref{eq:dyncon} holds
everywhere on the interface, except at the contact line, where the interface generally exhibits a kink and, accordingly, the curvature is undefined. A separate, pointwise condition must be specified at the contact line. We impose that the external traction is balanced by the the surface-tension tractions acting on both sides of the contact line, according to
\begin{equation}
\label{eq:Neumann}    
\gamma_{\textsc{lv}}\mathbf{t}_{\textsc{lv}}-\gamma_s(\mathbf{t}_-+\mathbf{t}_+)=0
\end{equation}
where $\mathbf{t}_{\pm}$ are the unit conormal vectors, normal to the contact line, and tangential and external to the smooth parts of the interface adjacent to the contact line. It is noteworthy that~\eqref{eq:Neumann} in fact corresponds to the Neumann law. The conditions (\ref{eq:dyncon}) and~(\ref{eq:Neumann}) are naturally satisfied in the equilibrium state \cite{pandey2020singular}, and imposing them dynamically ensures there is no dissipation associated to capillarity. Likewise, the adjacency of the liquid reservoir implies that the chemical potential at the interface vanishes at equilibrium, and we choose to impose $\mu=0$ dynamically to avoid any source of dissipation at the interface.  

The part of the substrate boundary complementary to the interface comprises the bottom and lateral boundaries. We assume that the poroelastic substrate is subjected to a preswelling treatment, in such a manner that the preswelling leads to a homogeneous isotropic expansion of the dry substrate with swelling ratio $\lambda$. Subsequently, the bottom boundary is regarded as fixed in the preswollen configuration and, accordingly, we impose~$\boldsymbol{\psi}(\mathbf{X},t)=\lambda\mathbf{X}$. On the lateral boundaries, we impose free-slip in the preswollen configuration, i.e. $\mathbf{n}\cdot\boldsymbol{\psi}(\mathbf{X},t)=\mathbf{n}\cdot\lambda\mathbf{X}$ and $\boldsymbol{n}\times(\boldsymbol{\sigma}\cdot \mathbf{n})\times\boldsymbol{n}=0$ with $\times$ the cross product. 
We regard the bottom and lateral boundaries as closed and therefore set $\boldsymbol{n}\cdot\nabla\mu=0$.

Our interest pertains to the evolution of the wetting ridge due to a point load at the interface associated with a liquid-vapor interface. We hence regard a preswollen initial configuration corresponding to the equilibrium of the poroelastic substrate in contact with the liquid reservoir, in the absence of the point load. The initial deformation then coincides with that of the preswollen configuration, i.e. 
$\boldsymbol{\psi}(\mathbf{X},t=0)=\lambda\mathbf{X}$,
and the initial concentration is uniform and such that the chemical potential (relative to the liquid reservoir) vanishes.



Thermodynamic consistency, in the sense of dissipation of the formulation without capillarity and without viscoelasticity, was discussed in \cite{hong2008theory}. The effect of $\boldsymbol \sigma_{\rm v}$ can be included in the energy-balance equation as well. Following the steps in \cite{pandey2020singular} and making use of the dynamical equations (\ref{eq:dcdt},\ref{eq:divsigma}) the free-energy functional evolves according~to 
\begin{equation}
\label{eq:dedt}
\mathrm{d}_t(\mathcal{E}-\mathcal{W})
=
- \int \mathrm{d}^2x\, \big(2\eta|\nabla \mathbf{v}|^2+D|\nabla \mu|^2\big)
 +  \operatorname{bnd}
\end{equation}
where $|\nabla \mathbf{v}|^2=\nabla \mathbf{v}:\nabla \mathbf{v}$ and $\operatorname{bnd}$ denotes terms on the part of the boundary of the substrate complementary to the interface; see Appendix~\ref{app:dissip}. Equation~\eqref{eq:dedt} conveys that the free energy is non-increasing, up to work supplied by liquid-vapor surface tension and, possibly, terms on the complementary part of the substrate boundary. Under the aforementioned boundary conditions, however, $\operatorname{bnd}$ vanishes so that dissipation only occurs in the bulk of the substrate. Moreover, because equilibrium configurations are characterized by vanishing dissipation, Equation~\eqref{eq:dedt} conveys that $\mathbf{v}$ and~$\mu$ are uniform in equilibrium. Subject to the aforementioned boundary conditions on the auxiliary part, $\mathbf{v}=0$ and $\mu=0$ in equilibrium.

\subsection{Weak formulation}
\label{sec:WeakForm}
%
%
To further elucidate the poro-visco-elastocapillary model, and to provide a basis for the finite-element approximation in section~\ref{sec:NumMeth}, we regard the weak formulation of~\eqref{eq:dcdt}--\eqref{eq:Neumann} constrained by~\eqref{eq:incompressibility}. 

A weak formulation of~\eqref{eq:divsigma} subject to the incompressibility constraint~\eqref{eq:incompressibility}, including the viscous part~\eqref{eq:viscoelastic}, with interface conditions~\eqref{eq:dyncon} and Neumann law~\eqref{eq:Neumann}, can be conveniently expressed in terms of derivatives of the energy and work functionals as
\begin{multline}
\label{eq:weakstruc}
(\boldsymbol{\psi}(\cdot,t),\Pi(\cdot,t)):
\\
\partial_{\boldsymbol{\psi}}\mathcal{E}[\boldsymbol{\psi},C,\Pi](\boldsymbol{W})
+\int\mathrm{d}^2x\,\nabla\mathbf{w}:\boldsymbol{\sigma}_{\text{v}}
\\
+\partial_{\Pi}\mathcal{E}[\boldsymbol{\psi},C,\Pi](V)=\partial_{\boldsymbol{\psi}}\mathcal{W}[\boldsymbol{\psi}](\boldsymbol{W})
\\
\quad\quad\forall(\mathbf{W},V)
\end{multline}
where $\partial_{\boldsymbol{\psi}}\mathcal{E}[\boldsymbol{\psi},C,\Pi](\boldsymbol{W})$ represents the Gateaux derivative of the free-energy functional with respect to its deformation argument in the direction~$\mathbf{W}$, in particular,
\begin{equation*}
\partial_{\boldsymbol{\psi}}\mathcal{E}[\boldsymbol{\psi},C,\Pi](\boldsymbol{W})
=\frac{\mathrm{d}}{\mathrm{d}s}\mathcal{E}[\boldsymbol{\psi}+s\boldsymbol{W},C,\Pi]\Big|_{s=0}
\end{equation*}
This notational convention for the Gateaux derivative carries over to other instantiations of $\partial_{(\cdot)}$. The functions $(\mathbf{W},V)$ act as test functions in the reference configuration, while $\mathbf{w}(\mathbf{x})=\mathbf{W}(\mathbf{X})$ is the representation of~$\mathbf{W}$ in the current configuration. The trial function $(\boldsymbol{\psi},\Pi)$ and test function~$(\mathbf{W},V)$ are supposed to belong to suitable admissibility classes, and $\boldsymbol{\psi}$ and~$\mathbf{W}$ are subject to conditions at the complementary boundary in accordance with imposed essential boundary conditions. By means of functional derivatives analogous to those in Appendix~\ref{app:dissip}, one can show that~\eqref{eq:weakstruc} is indeed equivalent to~\eqref{eq:divsigma}--\eqref{eq:Neumann}. 

The weak formulation of~\eqref{eq:dcdt} is most conveniently expressed in a form that represents the transport part in the reference configuration and the diffusive part in the current configuration:
\begin{equation}
C(\cdot,t):\quad
\frac{\mathrm{d}}{\mathrm{d}t}
\int\mathrm{d}^2X\,CY
+
\int\mathrm{d}^2x\,D\nabla{}\mu\cdot\nabla{}y=0
\quad\forall{}Y
\end{equation}
with $y(\mathbf{x})=Y(\mathbf{X})$.
The definition of the chemical potential in~\eqref{eq:chempot} is expressed in weak form as
\begin{equation}
\label{eq:weak_chempot}
M(\cdot,t):\quad
\int\mathrm{d}^2X\,MZ
-\partial_C\mathcal{E}[\boldsymbol{\psi},C,\Pi](Z)
=0
\quad\forall{}Z
\end{equation}
The weak formulation of the poro-visco-elastocapillary problem consists of the aggregation of~\eqref{eq:weakstruc}--\eqref{eq:weak_chempot}, equipped with suitable initial conditions for $\boldsymbol{\psi}$ and~$C$.






\subsection{Scales and dimensionless numbers}

The free-energy formulation provides the natural scales for energies and lengths. We start by comparing the Flory--Huggins energy density, scaling with $kT/v$, to the elastic energy density, scaling with $G=NkT$. Its ratio defines a first dimensionless parameters, $Nv$. Combined with the Flory--Huggins parameter $\chi$, this determines the natural degree of swelling of the substrate when fully immersed. These parameters are the natural ingredients of the Flory--Rehner theory \cite{flory1943statistical}.

Then, the ratio of surface energy to bulk elastic energy provides a length scale $\gamma_s/G$. This defines the so-called elasto-capillary length, which gives a typical measure of elastic deformation. Another length scale of the system is the layer thickness of the preswollen substrate, which we denote as $H$. The ratio $\gamma_s/(GH)$ geometrically describes the importance of the deformation. A final dimensionless parameter derived from the free energy is obtained from the ratio of surface tensions $\gamma_{\textsc{lv}}/\gamma_s$. This parameter is known to determine the solid angle of the solid, according to Neumann's law \cite{pandey2020singular,flapper2023reversal}.

In summary, there are 4 dimensionless numbers governing the ``statics" of the problem. For tractability, we keep them at fixed typical values, namely
\begin{equation}\label{eq:static}
Nv=1, \quad \chi=0, \quad \frac{\gamma_s}{GH}=1, \quad \frac{\gamma_{\textsc{lv}}}{\gamma_s}=1.
\end{equation}
Changing the parameter values will lead only to quantitative differences, in particular the degree of preswelling ($Nv$, $\chi$), the relative size of the elastic deformation ($\gamma_s/GH$), and the opening angle of the ridge ($\gamma_{\textsc{lv}}/\gamma_s$). However, our main interest here is to explore the dynamical behavior of the ridge, which is not expected to change qualitatively with these parameters.

The dynamical model is defined by (\ref{eq:dcdt}) and (\ref{eq:divsigma}), each of which introduces a natural timescale. The equation for solvent transport relates $c$ to $\mu$, which have typical scales $1/v$ and $kT \sim Gv$. To estimate the diffusion time, one can use either $\gamma_s/G$ (the ridge size) or $H$ (the substrate thickness) as the relevant length scale. We opt for the former, which gives the diffusion timescale inside the ridge,
\begin{equation}
\tau_{\rm D} = \frac{(\gamma_s/G)^2}{D v^2 G} = \frac{\gamma_s^2}{D v^2 G^3}.
\end{equation}
In the presence of viscous effects per~\eqref{eq:viscoelastic}, the momentum equation contains a visco-elastic relaxation timescale $\tau_{\rm VE} = \eta/G$, introducing a fifth dimensionless number
\begin{equation}
\tilde{\tau}=\frac{\tau_{\textsc{ve}}}{\tau_D} = \frac{\tau_{\textsc{ve}} D v^2 G^3}{\gamma_s^2}.
\end{equation}
This parameter governs the cross-over from poroelastic to viscoelastic response.

In the remainder of the paper we use tildes to denote dimensionless variables. We scale lengths with $\gamma_s/G$, energies with $vG$, while time is scaled with $\tau_D$. Hence, results will be presented in terms of the following dimensionless variables: 

\begin{equation}
\tilde{\mathbf{x}}=\frac{\mathbf{x}}{\gamma_s/G}, \quad 
\tilde{t} = \frac{t}{\tau_D}, \quad
\tilde{\mu}=\frac{\mu}{vG},  \quad
\tilde{c}=v c.
\end{equation}
Note that the solvent concentration is made dimensionless with molecular volume (rather than $\gamma_s/G$), so that $\tilde c$ directly gives the liquid volume fraction.




\subsection{Numerical methods}
\label{sec:NumMeth}
We approximate the poro-visco-elastocapillary problem~\eqref{eq:weakstruc}--\eqref{eq:weak_chempot} by means of the finite-element method in the spatial dependence, and the implicit Euler method in the temporal dependence. The approximation method has been implemented in the open-source software framework Nutils~\cite{nutils70}.

To provide adequate resolution at the tip of the wetting ridge, we apply a mesh with a-priori local refinements near the contact line. Starting with a coarse baseline mesh, we successively subdivide the elements contained in a semi-disc with increasingly small radius, centered at the contact line; 
see Figure~\ref{fig:epsart}(c). The discrete approximation is obtained by replacing the ambient spaces for the trial functions $\boldsymbol{\psi},\Pi,C,\mu$ and the test functions~$\boldsymbol{W},V,Y,Z$ by finite-element spaces subordinate to the locally refined mesh. In particular, we apply continuous piece-wise quadratic approximation spaces for $\boldsymbol{\psi},C,\mu$ and~$\boldsymbol{W},Y,Z$ and continuous piece-wise linear approximation spaces for~$\Pi$ and~$V$. This implies that for the deformation/pressure pair $(\boldsymbol{\psi},\Pi)$ we apply so-called Taylor--Hood elements. The functional derivatives that appear in the weak formulation~\eqref{eq:weakstruc}--\eqref{eq:weak_chempot} are obtained from implementations of the free-energy and work functionals, by means of the automatic differentiation functionality in Nutils.

The nonlinear algebraic problem that emerges in each time step, is solved by means of a Newton--Raphson method with line search. The linear tangent problems that occur in each iteration of the Newton--Raphson method are solved by a direct solver.  We validate the numerical results by comparing the interface profile at $ \tilde{t} = 0^+ $ and $\tilde{t} = \infty$   with the equilibrium results presented in \cite{pandey2020singular, flapper2023reversal}. Further details are provided in Appendix A.




\section{\label{sec:level1}Poroelasticity}

\begin{figure*}[ht]
\includegraphics[width=1\textwidth]{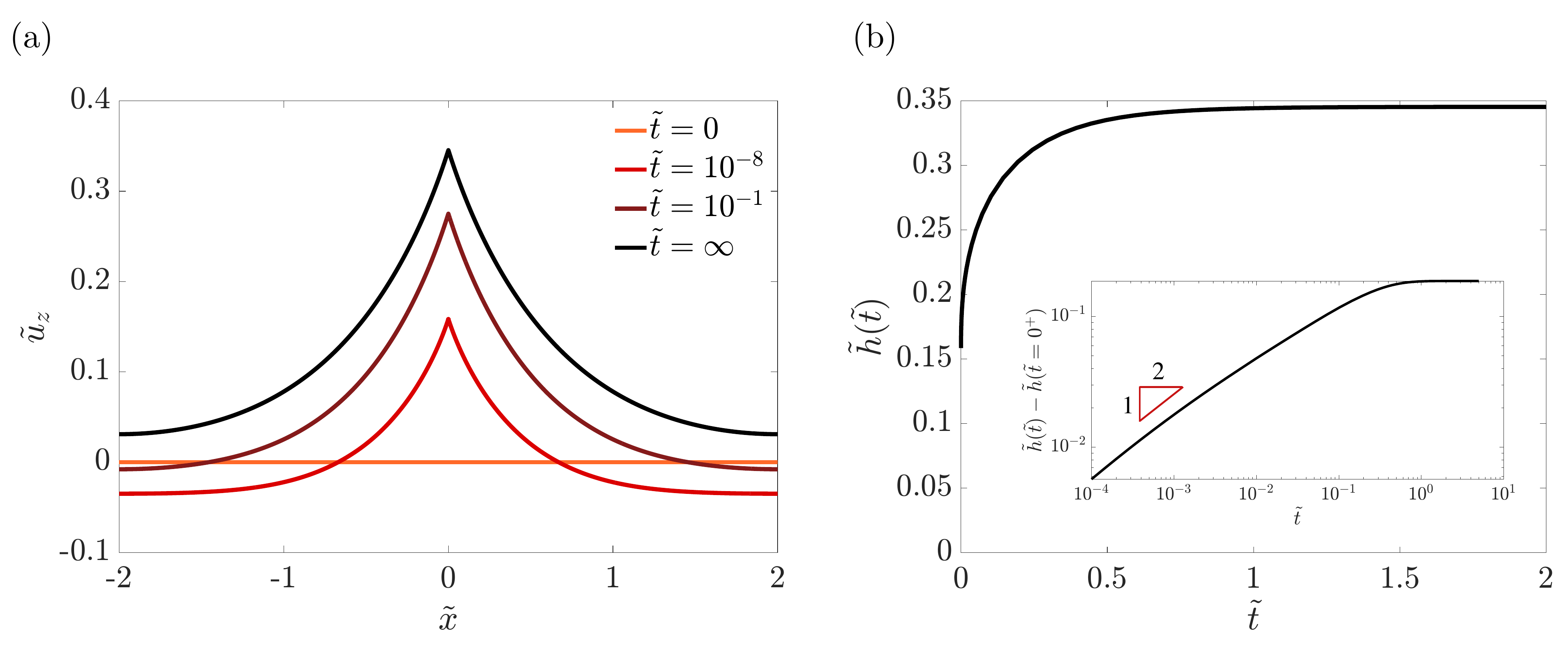}
\caption{\label{fig:porouz} (a) Time-dependent profiles of the free surface of the purely poroelastic substrate ($\tilde \tau=0$) after the application of a point force at $\tilde t=0$; (b) The ridge height $\tilde h$ as a function of time. Inset: Evolution of the ridge height $\tilde{h}(\tilde{t}) - \tilde{h}(0^+)$ after the instantaneous response, on a logarithmic scale.}
\end{figure*}

\begin{figure*}[ht]
\includegraphics[width=1\textwidth]{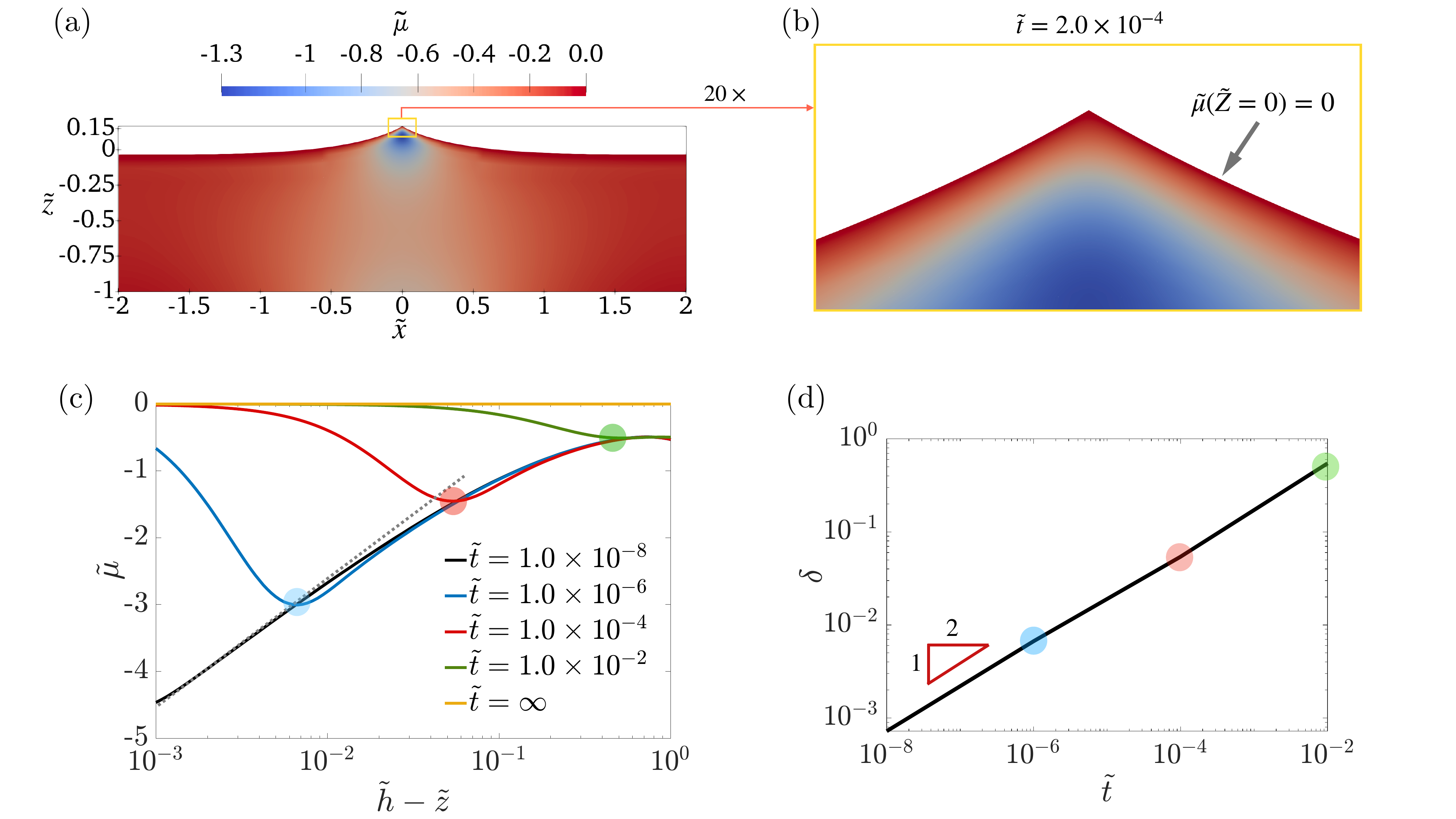}
\caption{\label{fig:porochem} Spatio-temporal evolution of the chemical potential for a porelastic substrate ($\tilde \tau=0$). (a) The rescaled chemical potential field at $\tilde{t} = 2 \times 10^{-4}$; (b) A zoomed-in view near the contact line position from (a); (c) Measurement of $\tilde{\mu}$ from the tip along the $z$-direction at different times. The circles indicating the local minima define the boundary layer thickness $\delta$. The dashed line indicates the prediction (\ref{eq:mulog}) for the instantaneous incompressible response; (d) Boundary layer thickness $\delta$ as a function of time $\tilde{t}$.}
\end{figure*}

We first present results for the purely poroelastic model, assuming there is no viscoelastic dissipation. This case corresponds to $\tilde{\tau}=0$, so that the polymer network adapts instantaneously to the imbibition of the solvent without any effects of viscoelastic relaxation. 
We recall that we consider the gel to be initially preswollen, equilibrated with the reservoir. For the parameters defined by (\ref{eq:static}), this implies the initial swelling ratio $J_0=1.582$. At $\tilde t=0$ we impose a contact line force at the free surface, located at $(\tilde X,\tilde Z)=(0,0)$. Since the contact line force pulls vertically upwards, the contact line position remains at $\tilde x=0$ in the deformed configuration. 

\subsection{Ridge growth}

To illustrate the evolution of the free surface of the poroelastic gel, Figure~\ref{fig:porouz}(a) plots the vertical displacement of the free surface, $\tilde{u}_z(\tilde{x},\tilde{t})$, versus the horizontal position in the deformed configuration, $\tilde{x}$, at various instants, $\tilde{t}$. At $\tilde t = 0$, the polymeric gel surface is flat, corresponding to the homogeneously preswollen initial state. Upon application of the capillary force at the contact line, the gel exhibits an instantaneous response, as evidenced by the profile at $\tilde{t} = 10^{-8}$. We refer to this very early time limit as $\tilde{t} = 0^+$. This instantaneous response is essentially incompressible because the solvent within the gel did not yet have any time to diffuse, preventing any noticeable change in swelling immediately after the capillary force is applied; see also~\cite{ZhaoMenghua2018}. Subsequently, as time progresses, the gel absorbs solvent from the surroundings and swells, resulting in an increase in its volume, as indicated by the curve at $\tilde{t} = 10^{-1}$ and the profile at $\tilde{t}=\infty$, in Fig.~\ref{fig:porouz}(a). The substrate eventually reaches an equilibrium state as $\tilde{t}\rightarrow \infty$. 

To further characterize the dynamics of the ridge growth, we show the height of the elastocapillary ridge, 
$ \tilde{h}(\tilde{t}):=\tilde{u}_z(0,\tilde{t})$, as a function of time in Fig.~\ref{fig:porouz}(b).
We indeed observe an instantaneous displacement at $\tilde{t}=0^+$, which is followed by a gradual increase until equilibrium is reached. The inset of Fig.~\ref{fig:porouz}(b) presents the same data, but instead plots $\tilde{h}(\tilde{t}) - \tilde{h}(0^+)$ 
versus~$\tilde{t}$ on a logarithmic scale. This graph shows that after the instantaneous incompressible response, the growth of the ridge follows a power law with an exponent of 1/2. In dimensional terms, this growth implies that $h(t) - h(0^+)\sim (D v^2 G t)^{1/2}$, which is scale-free in the sense that it does not involve the elasto-capillary length $\gamma_s/G$ nor the system size $H$. This scale-free behavior reflects that the growth is initially due to diffusive imbibition of the solvent into the substrate. After this scaling regime, the equilibration is observed around $\tilde t \sim 1$, suggesting that the diffusion timescale $\tau_D$, is indeed the appropriate poroelastic timescale for the late-time process, where the ridge size $\gamma_s/G$ begins to play a role. 

\subsection{Chemical potential}

Solvent transport plays a fundamental role in the dynamics of the wetting-ridge formation. The chemical potential~$\mu$ per~(\ref{eq:chempot}) effectively controls the magnitude and direction of the solvent flux, because the flux is proportional 
to~$-\nabla \mu$. Hence, the spatio-temporal evolution of~$\mu$ provides meaningful insight into the underlying solvent-transport mechanisms. 

Figure~\ref{fig:porochem}(a) presents the rescaled chemical 
potential~$\tilde \mu$ in the porous substrate, at $ \tilde{t} = 2 \times 10^{-4} $. The figure conveys that the chemical potential displays a localized minimum just below the contact line, represented in the figure by  the blue region. This minimum in the chemical potential can equivalently be interpreted as a region of low ``pore pressure", and originates from the tensile elastic stress generated by the deformed polymer network. The solvent thus tends to diffuse towards this blue region. A magnified view near the contact line is given in Fig.~\ref{fig:porochem}(b). This zoom shows that $\tilde \mu=0$ at the free surface, reflecting the boundary condition that the surface is in equilibrium with the reservoir. The blue region is separated from the free surface by a thin boundary layer, that gives rise to a strong local gradient of $\tilde \mu$. Hence, the solvent flux will be strongest near the interface, so that the predominant flux is from the reservoir into the ridge.


Next, we focus on the temporal evolution of the chemical potential. Figure~\ref{fig:porochem}(c) shows $\tilde \mu$ along the vertical line below the contact line ($\tilde x$=0), at different times. The horizontal axis $\tilde h - \tilde z$ represents the distance to the contact line. The graphs confirm the appearance of a minimum of $\tilde \mu$, which shifts and fades over time. At very long times $\tilde \mu \to 0$ everywhere, corresponding to the global equilibration. Interestingly, the data for early times ($\tilde{t} = 10^{-8}$, shown in black) seem to follow a logarithmic dependence $\tilde \mu \sim \ln(\tilde h - \tilde z)$. This logarithmic behavior can be understood in terms of the instantaneous incompressible response expected at $t=0^+$. At this early-time limit, the concentration field did not yet have time to evolve, and the chemical potential, as defined in (\ref{eq:chempot}), follows the behavior of $\Pi$. For an incompressible corner, $\Pi$ plays the role of the pressure that is known to exhibit a logarithmic dependence on the distance to the contact line \cite{pandey2020singular}. Translated in terms of the scaled chemical potential, this takes the form
\begin{equation}\label{eq:mulog}
\tilde{\mu} \simeq \left(\frac{\pi}{\theta_{\textsc{s}}}-\frac{\theta_{\textsc{s}}}{\pi}\right) \ln ( \tilde r),
\end{equation}
where $\tilde r$ is the distance to the contact line and $\theta_{\textsc{s}}$ is the corner angle ($\theta_{\textsc{s}}=2\pi/3$ when  $\gamma_{\textsc{lv}}/\gamma_s=1$). 
This prediction is superimposed as the gray dashed line in Fig.~\ref{fig:porochem}(c), and indeed perfectly captures the observed behavior in the vicinity of the contact line at $\tilde{t}=10^{-8}$.

The picture that emerges is that the instantaneous response dictates the chemical potential to diverge logarithmically, according to (\ref{eq:mulog}), in the vicinity of the contact line. This behavior, however, is incompatible with the boundary condition $\mu=0$ that is imposed at the free surface. This conundrum is resolved by the emergence of a thin boundary layer near the contact line. Indeed, the blue and red curves in Fig.~\ref{fig:porochem}(c) closely follow the near-instantaneous response (black line) away from the contact line. At shorter distances, the logarithmic trend is not followed, and the data tend to $\tilde \mu=0$ at the contact line. We estimate the size of the boundary layer $\delta$ by tracing the minimum of the chemical potential, indicated by the circles. Figure~\ref{fig:porochem}(d) reports the boundary layer thickness over time, which again follows a diffusive $1/2$ scaling. Hence, we associate the dynamics of the boundary layer with the scale-free diffusive solvent transport.

\subsection{Swelling dynamics}

\begin{figure*}[ht]
\includegraphics[width=1\textwidth]{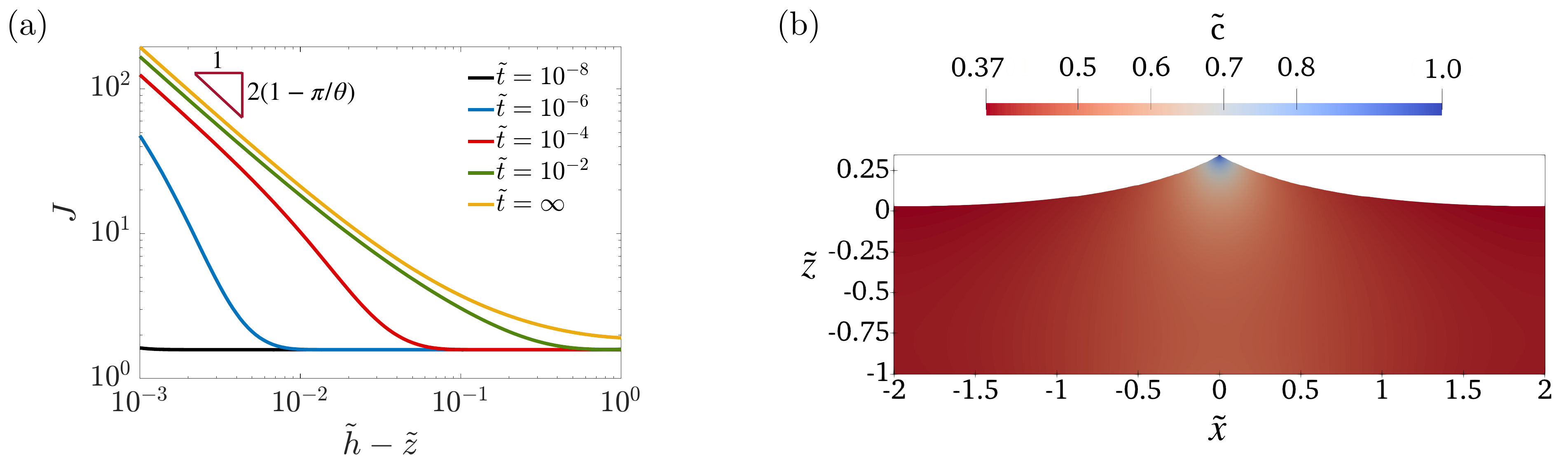}
\caption{\label{fig:poroc} Swelling dynamics for a porelastic substrate ($\tilde \tau=0$). (a) Measurement of $J$ from the tip along the $z$-direction at different times. The preswelling at $t=0$ corressponds to an initial $J_0=1.582$. (b) Solvent volume fraction field $\tilde c$ at the final equilibrium state. Near the tip the solvent fraction $\tilde c \to 1$, while at large distance one recover the value due to preswelling $\tilde c \approx 0.37$. 
}
\end{figure*}

From the previous results we thus conclude that the incompressible response gives an excellent representation of the behavior at early times, except for a narrow boundary layer $\delta \sim t^{1/2}$. Swelling does occcur in this boundary layer, which is characterized by a strong flux $\sim 1/\delta \sim 1/t^{1/2}$. Now we turn to the question of how this translates to the spatio-temporal dynamics of the swelling inside the substrate. 

Figure~\ref{fig:poroc}(a) shows the swelling ratio $J$ as a function of the vertical distance $\tilde h - \tilde z$, measured directly below the contact line. The initial value corresponds to the preswollen substrate with $J_0=1.582$. In line with the evolution of the boundary layer for $\tilde \mu$, we observe that swelling is initiated close to the contact line, and gradually invades the entire domain. The long-time equilibrium profile is given by the yellow line. The strongest swelling  occurs at the tip of the wetting ridge, and exhibits a power-law singularity. Indeed, previous work on the equilibrium state \cite{flapper2023reversal} has analytically demonstrated that the swelling ratio $J \sim \tilde r^\beta$, where the exponent $\beta = 2(1 - \frac{\pi}{\theta_{\textsc{s}}})$; in this expression $\theta_{\textsc{s}}$ once again is the opening angle of the tip of the solid. The numerical results closely follow this analytical prediction, also transiently in the vicinity of the contact line. 

The divergence of $J$ implies that the tip-region is almost purely liquid. Specifically, the scaled concentration $\tilde c = 1 - 1/J$ directly gives the liquid volume fraction. The volume fraction at equilibrium is shown in Fig.~\ref{fig:poroc}(b), and indeed approaches unity near the tip. In the present formulation the liquid cannot phase-separate between the matrix and the liquid reservoir, as observed in some experiments \cite{jensen2015wetting, cai2021fluid,hauer2023phase}. It is clear, however, that the strong aspiration of liquid in elastic corners provides a natural mechanism for this phenomenon.

\section{\label{sec:level1} Visco-poroelasticity}

\begin{figure}
\includegraphics[width=0.48\textwidth]{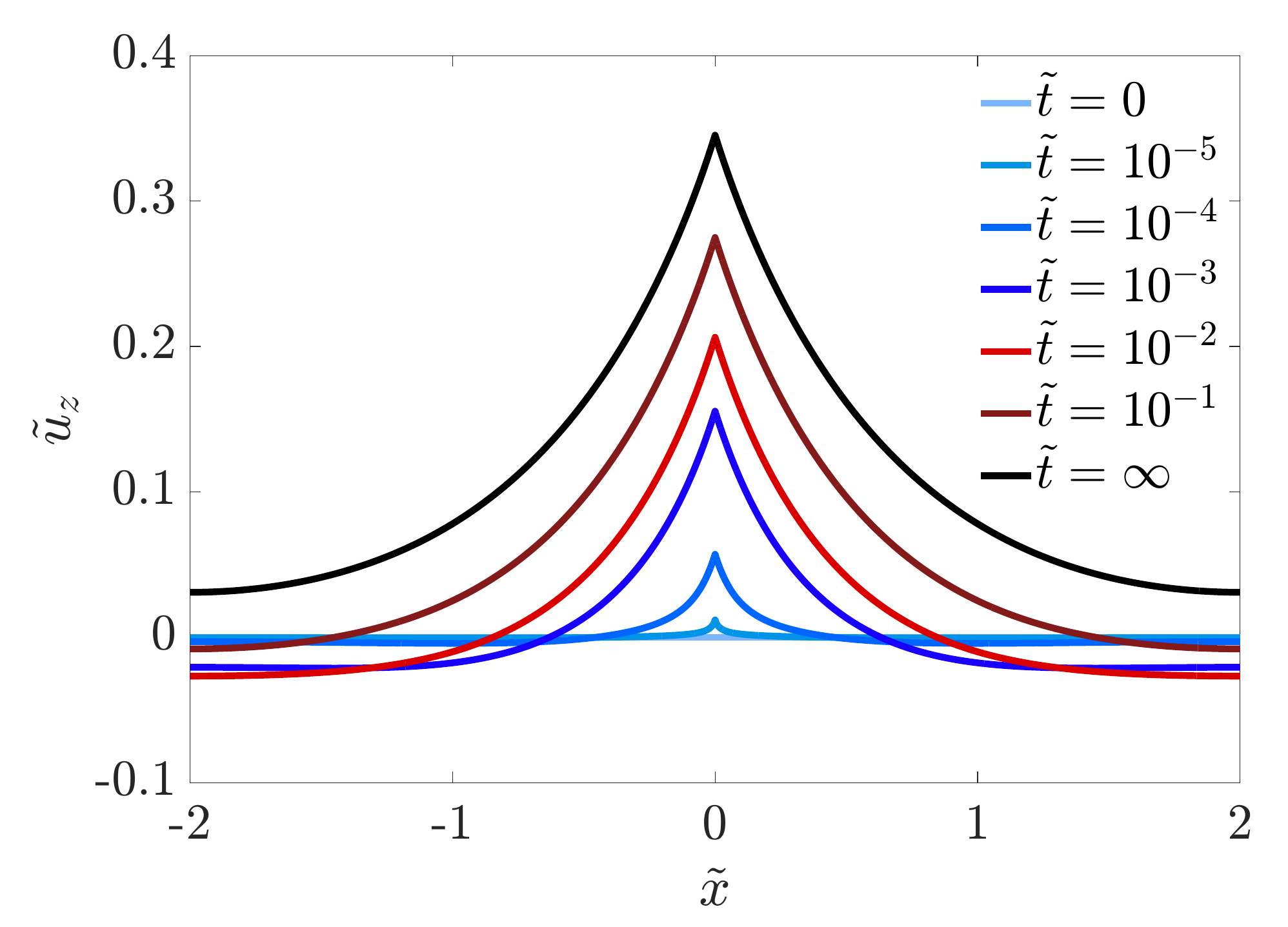}%
\caption{\label{fig:viscoprofile} Time-dependent surface profile for a visco-poroelastic substrate ($\tilde \tau= 10^{-3}$), after the application of a point force at $\tilde t=0$. For $\tilde t < \tilde \tau$, a small ridge emerges gradually from the substrate with a viscoelastic dynamics. For $\tilde t > \tilde \tau$, the profiles closely follow the purely poroelastic dynamics of Fig.~\ref{fig:porouz}(a).}
\end{figure}

We now turn to the complete model, including the effect of viscoelasticity. This case corresponds to $\tilde{\tau}= \tau_{\rm VE}/\tau_D \neq 0$, for which there is a competition of  two timescales: the poroelastic diffusive timescale $\tau_D$ and the viscoelastic timescale $\tau_{\rm VE}$ reflecting the relaxation of the polymer network. 

\subsection{Ridge growth}

Figure~\ref{fig:viscoprofile} depicts a typical visco-poroelastic evolution of the free surface for the case where $\tilde \tau=10^{-3}$. This value of $\tilde \tau$ implies that the network relaxation is faster compared to the diffusive transport of the solvent. Nonetheless, we observe that the growth of the ridge is fundamentally altered with respect to the purely poroelastic case: the instantaneous incompressible response is now replaced by a gradual viscoelastic growth of the wetting ridge, occurring on the timescale $\tilde t < \tilde \tau$. The ridge emerges gradually in this regime, both in terms of its height and its width. It is noteworthy that the solid angle of the ridge forms instantaneously, 
in accordance with the Neumann law~\eqref{eq:Neumann}, which is implicit in~\eqref{eq:weakstruc} as a natural (weakly imposed) condition; see also~\cite{karpitschka2015droplets,Chan2022}.  We note that after $\tilde t \approx \tilde \tau =10^{-3}$ (dark blue), the profiles closely follow those in Fig.~\ref{fig:porouz}(a). The limiting transport mechanism during this late-time dynamics is the slow poroelastic solvent flow, not the viscoelastic relaxation.

\begin{figure*}[ht]
\includegraphics[width=1\textwidth]{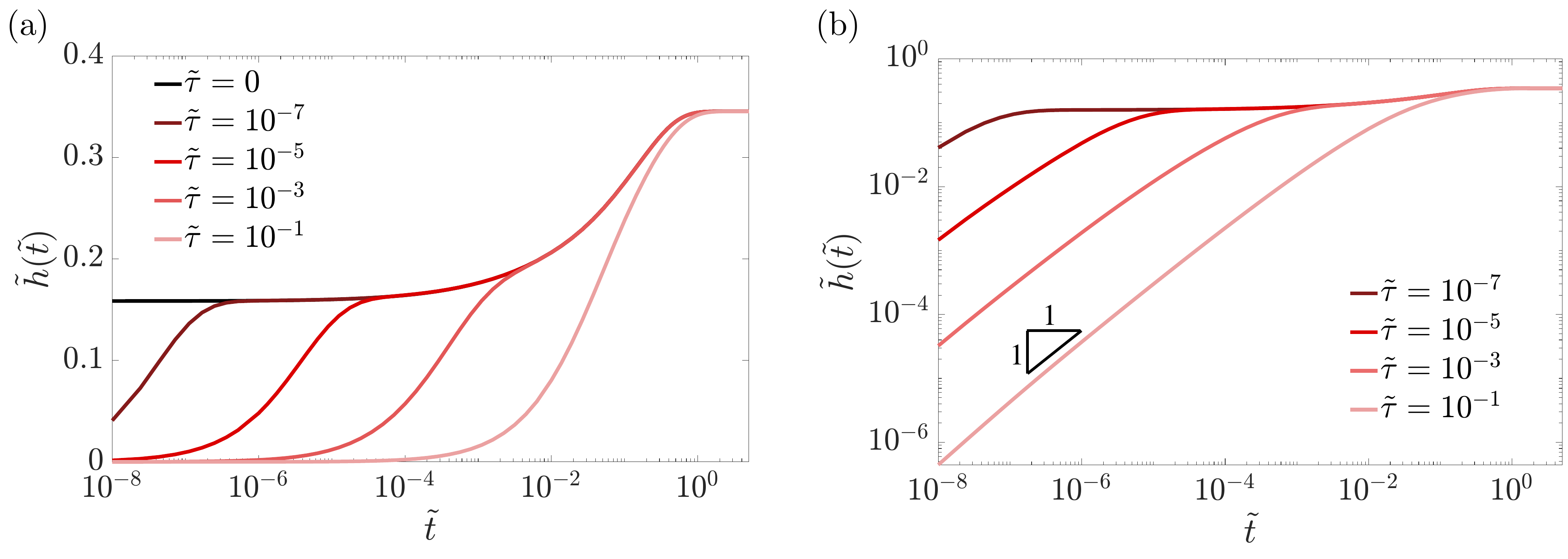}
\caption{\label{fig:vispouz}  Growth of the ridge for visco-poroelastic substrates. (a) $\tilde{h}$ The deformation of the tip as a function of time $\tilde{t}$ for various viscoelastic relaxation times $\tilde{\tau}$. (b) Same data on a double logarithmic scale, showing the linear ridge growth when viscoelasticity is included.}
\end{figure*}

To further quantify the ridge dynamics, we report in Fig.~\ref{fig:vispouz}(a) the height of the ridge $\tilde{h}$ as a function of time $\tilde{t}$ in a semi-log plot, for various viscoelastic relaxation times $\tilde{\tau}$. When $\tilde{\tau}=  0$ (represented by the black line), we recover the purely poroelastic case that we previously observed to exhibit an instantanuous response. Conversely, when $\tilde{\tau}$ is non-zero, one no longer observes this instantaneous deformation. Instead, the tip height grows gradually until joining the purely poroelastic curve approximately when $\tilde t \sim \tilde \tau$. These results confirm the crossover, from viscoelastic at early times to  poroelastic at later times. Figure~\ref{fig:vispouz}(b) shows the same data on a doubly logarithmic plot. We find a linear initial growth in the viscoelastic regime. The scaling follows $\tilde h \sim \tilde t/\tilde \tau$, which in dimensional units amounts to $h \sim \gamma_s t/(G\tau_{\rm VE})$. Bearing in mind that the network viscosity $\eta = G \tau_{\rm VE}$, the growth of the ridge is dictated by the capillary velocity $\gamma_s/\eta$.

\subsection{Swelling dynamics}

\begin{figure*}[ht]
\includegraphics[width=1\textwidth]{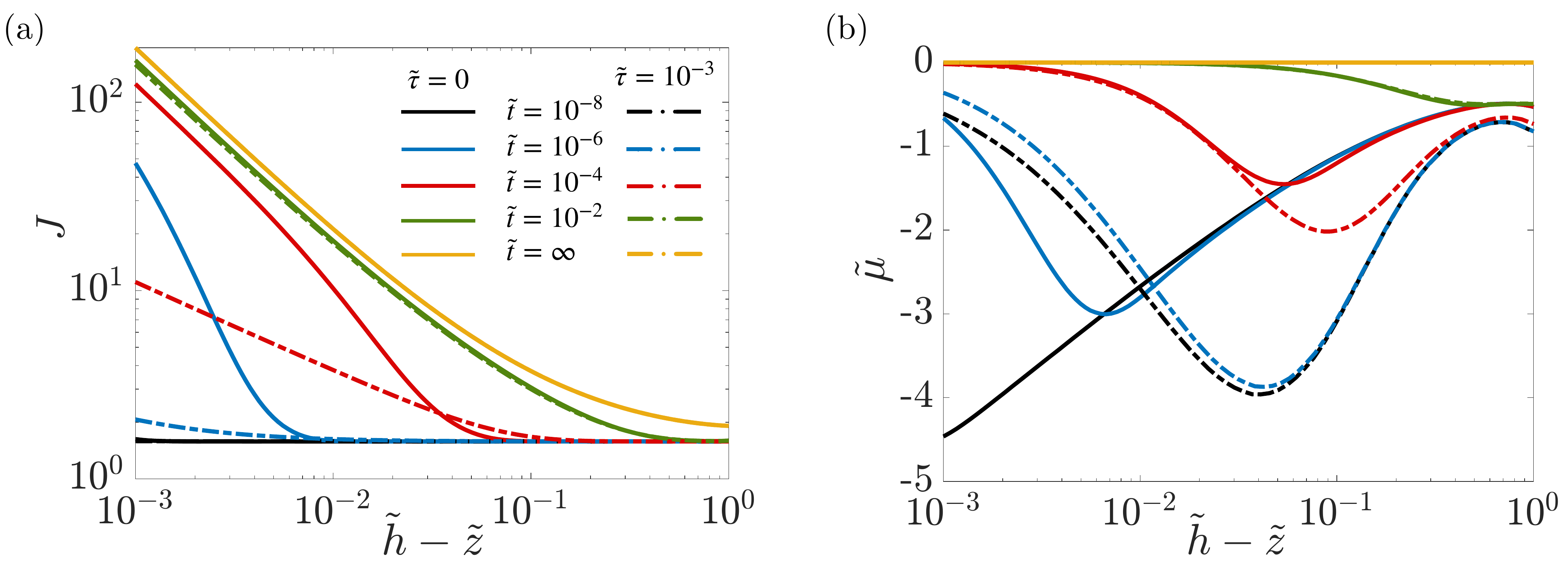}
\caption{\label{fig:vispoin} Swelling ratio (a) and chemical potential (b), for a visco-poroelastic substrate. Profiles along the $z$-direction from the tip. The solid lines corresponds to a purely poroelastic material ($
\tilde \tau=0$). The dashed lines correspond to a visco-poroelastic material with $\tilde{\tau} = 10^{-3}$.}
\end{figure*}

Finally, we turn to the transport of solvent in the presence of viscoelasticity. Figure~\ref{fig:vispoin}(a) shows the swelling ratio below the contact line, at different times. The dash-dotted lines represent the results for the case where $\tilde{\tau} =  10^{-3}$, which was previously considered in Fig.~\ref{fig:viscoprofile}. To clearly illustrate how viscoelasticity changes the transport dynamics, we  also report the purely poroelastic results for $\tilde{\tau} = 0$, as  solid lines. For times $\tilde t > \tilde \tau$, the solid lines and dash-dotted lines essentially overlap. This implies that the late-time swelling dynamics is not affected by viscoelasticity, but is purely poroelastic in nature. By contrast, large differences are observed for the early time dynamics, $\tilde t < \tilde \tau$. There is still a tendency towards swelling near the tip, even at very short times, but this swelling is retarded compared to the purely poroelastic case.  The suppression of solvent transport at early times can be attributed to the vanishing of logarithmic divergence in the chemical potential, as  shown in Figure~\ref{fig:vispoin}(b). In other words, the viscoelastic stresses inhibit the development of strong ``pore pressure''. Only when $\tilde t \geq \tilde \tau$, the curves for $\tilde \tau =0 $ and $\tilde \tau \neq 0$ start to overlap, indicating the regime where transport is purely poroelastic.

\section{\label{sec:level1}Conclusion and Remarks}

In this study, we investigate the deformation and solvent transport within a poro-viscoelastic polymeric gel under the influence of capillary forces. A large deformation model is employed to accurately characterize the polymeric gel’s deformation, while the Flory--Huggins theory is utilized to describe the mixing energy between the solvent and the polymeric gel. We assume a non-inertial motion for the system and incorporate viscoelasticity to describe the polymeric gel's complex behavior.

We begin by examining purely poroelastic substrates, where viscoelastic effects are absent. When a capillary force is applied at the contact line, the substrate responds instantaneously as an incompressible material, resulting in immediate deformation at $\tilde{t} = 0^+$. Subsequently, the substrate exhibits poroelastic behavior, reflected in the swelling and the growth of the ridge. The increase in the tip's height over time follows a power law with an exponent of 1/2, implying that the early-time dynamics are scale-free and independent of the elastocapillary length. This result appears to be at variance with the study by Zhao et al. \cite{ZhaoMenghua2018}, who investigate the ridge growth for an axisymmetric droplet using a linear elasticity theory. The authors report an instantaneous response, similar to our observations, but they describe the subsequent dynamics as following a logarithmic relation in the time range $\tau_D\ll t\ll (RG/\gamma_s)^2\tau_D$, where $R$ is the droplet radius.  We suspect the dynamics in~\cite{ZhaoMenghua2018} are, in fact, governed by a $t^{1/2}$ scaling at very early times, with the logarithmic behavior emerging as an intermediate asymptotic regime specific to the axisymmetric case.

Regarding solvent transport, we observe a logarithmic singularity in the chemical potential at the tip at $\tilde{t} = 0^+$.
This singularity is regularized by a thin boundary layer, characterized by a large solvent flux from the adjacent liquid reservoir into the substrate near the tip. Over time, the minimum in chemical potential is found to move away from the tip into the substrate, driving the gradual swelling of the ridge into the substrate.  Additionally, the swelling ratio exhibits a singularity at the contact line, indicating that the volume fraction of the solvent approaches unity near the tip. The relationship between the swelling ratio and the distance from the tip can be described by a power law with an exponent of  $2(1 - \pi/\theta_{\textsc{s}})$, in line with previous predictions for the equilibrium case \cite{flapper2023reversal}. This contrasts previous results for small deformations \cite{ZhaoMenghua2018}, for which a logarithmic divergence of the swelling ratio has been predicted.

The incorporation of viscoelastic effects in the substrate reveals that viscoelasticity dominates the early time dynamics of the ridge growth (prior to the relaxation time) and significantly suppresses  solvent transport. Beyond the viscoelastic relaxation time, the dynamics converge to those of the purely poroelastic case.

Finally, let us discuss the interpretation of recent experiments in light of our findings. 
First, we remark that our results are consistent with the experimental results of Xu et al. \cite{xu2020viscoelastic}, which examined the relaxation of an equilibrium ridge. Specifically, the early-time dynamics are dominated by viscoelastic effects, while poroelasticity becomes the primary factor governing the late-time behavior. 
However, the reversed scenario was proposed in \cite{li2023crossover}, who measured the force relaxation using Atomic Force Microscopy after pressing an indenter into the hydrogel. These experiments exhibited a slow decay of the measured force at long times, which was attributed to viscoelasticity. These experiments were conducted on agarose gels, which might be susceptible to plastic reconfigurations that manifest themselves as long-time creep. Clearly, this kind of material would require a different class of viscoelastic models compared to what was used in our simulations. When staying within the realm of linear viscoelasticity, a long-time relaxation scenario would require a vanishing stress relaxation function, with no resultant elasticity at large times. In addition to these conclusions, our theoretical results reveal the mechanism by which solvent is transported to the wetting ridge. This offers a route that could explain the experimentally observed extraction of solvent at the contact line \cite{cai2021fluid}, and shows the potential of the presented modeling approach.

\begin{acknowledgments}
 BXZ and TSC gratefully acknowledge the financial support from the Research Council of Norway (project no. 315110). JHS acknowledges support from NWO through VICI grant no. 680-47-632. BXZ would also like to extend heartfelt gratitude to Dr. Gertjan van Zwieten and Dr. Martin H. Essink for their valuable insights and assistance with coding. 

\end{acknowledgments}

\appendix

\section{Validation}\label{app:validation}

\begin{figure}
\includegraphics[width=0.5\textwidth]{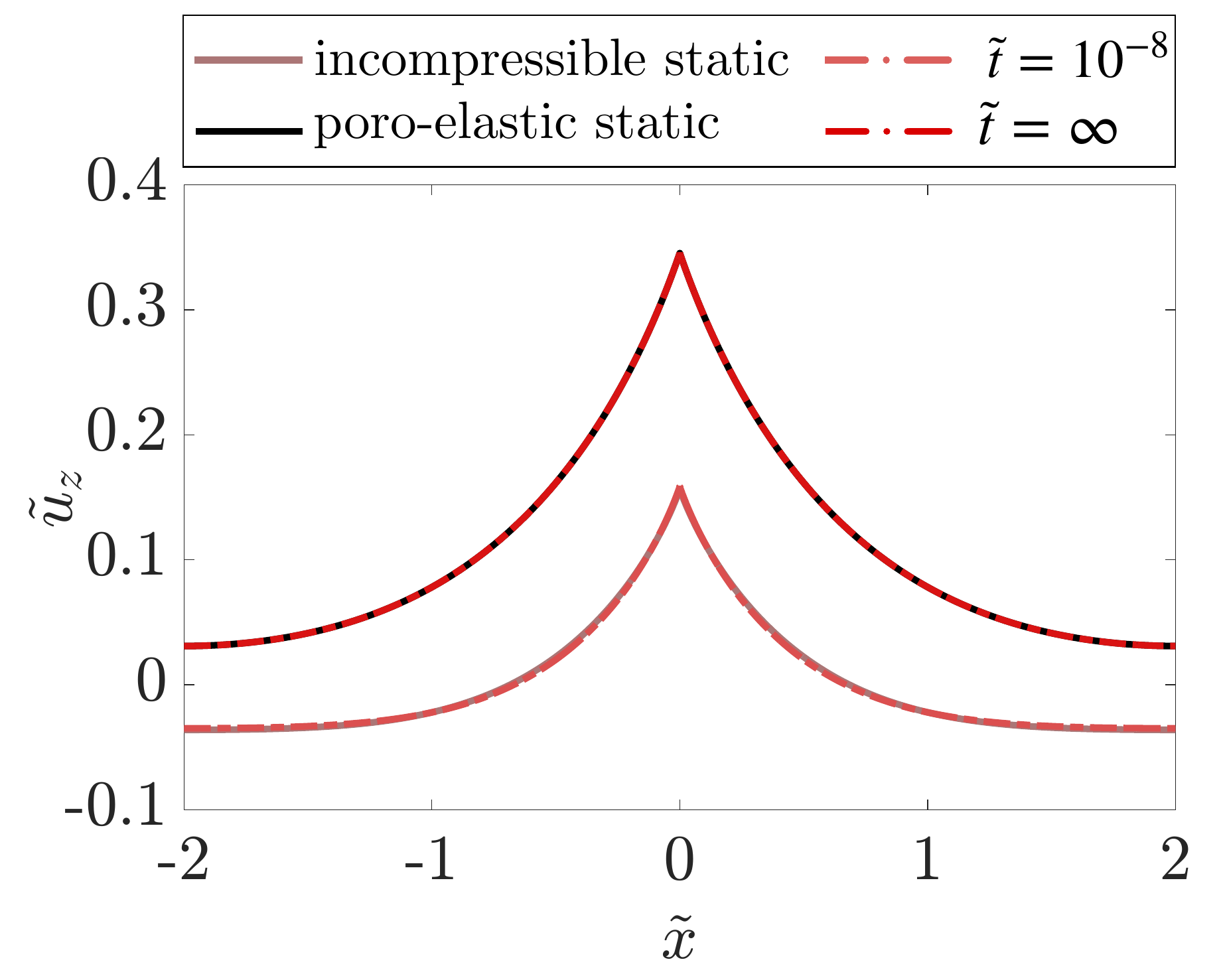}
\caption{\label{fig:appendix} Comparison of dynamic and static wettability profiles on the polymeric gel. The profiles at early time ($\tilde{t} = 10^{-8}$) and late time ($\tilde{t} = \infty$, equilibrium) are compared with static wettability results for incompressible and poroelastic materials, respectively.}
\end{figure}

We validate our computations by comparing the dynamic results obtained in this study with previous static findings related to wetting on incompressible and poroelastic substrates. As analyzed previously, at $ t = 0^+$, the substrate behaves as an incompressible material. Given that we are considering quasi-static motion in the present case, the deformation of the substrate at $ t = 0^+$ should align with the static wetting behavior observed on an incompressible substrate. In the equilibrium state ($ t = \infty$), the substrate profile is expected to converge with the static wetting results for a poroelastic material. The profiles of the substrate surface from prior studies on wetting in incompressible substrates \cite{pandey2020singular} and poroelastic substrates \cite{flapper2023reversal} have been plotted and compared with the present results for dynamic wetting on the poroelastic substrate in Fig.~\ref{fig:appendix}. The agreement serves as a validition of our time-dependent numerical solutions.

\section{Dissipation}
\label{app:dissip}
To derive the dissipation relation~\eqref{eq:dedt}, we first note that
\begin{equation}
\label{eq:dissip1}
\begin{aligned}
\mathrm{d}_t\mathcal{E}[\boldsymbol{\psi},C,\Pi]&=
\int \mathrm{d}^2X\,(\mathrm{d}_{\mathbf{F}}W_{\text{el}}-\Pi\,\mathrm{d}_{\mathbf{F}}J):\partial_t\mathbf{F}
\\
&\phantom{=}
+
\int \mathrm{d}^2X\,
(W_{\textsc{fh}}'(C)+\Pi{}v)\partial_tC
\\
&\phantom{=}
+
\int \mathrm{d}^2X\,
(1+vC-J)\,\partial_t\Pi
+
\mathrm{d}_t\int{}\mathrm{d}s\,\gamma_s
\end{aligned}
\end{equation}
By virtue of the static equilibrium condition and the partition $\boldsymbol{\sigma}=\boldsymbol{\sigma}_{\text{el}}+\boldsymbol{\sigma}_{\text{v}}$, 
via standard transformations and integration-by-parts identities we obtain for 
the first term: 
\begin{equation}
\begin{aligned}
&\int \mathrm{d}^2X\,(\mathrm{d}_{\mathbf{F}}W_{\text{el}}-\Pi\,\mathrm{d}_{\mathbf{F}}J):\partial_t\mathbf{F}
\\
&\quad=
\int \mathrm{d}^2x\,\boldsymbol{\sigma}_{\text{el}}:\nabla\mathbf{v}
+
\int \mathrm{d}^2x\,\mathbf{v}\cdot(\nabla\cdot\boldsymbol{\sigma})
\\
&\quad=
\int\mathrm{d}s\,\mathbf{v}\cdot\boldsymbol{\sigma}\cdot \mathbf{n}-
\int\mathrm{d}^2x\,\boldsymbol{\sigma}_{\mathrm{v}}:\nabla\mathbf{v}
\end{aligned}    
\end{equation}
Note that the second term in the central expression in fact vanishes per~\eqref{eq:divsigma}.
For the second term in~\eqref{eq:dissip1}, it follows from~\eqref{eq:dcdt}, the Reynolds transport theorem and the divergence theorem that
\begin{equation}
\begin{aligned}
&\int \mathrm{d}^2X\,
(W_{\textsc{fh}}'(C)+\Pi{}v)\,\partial_tC
=
\int \mathrm{d}^2X\,M\partial_tC
\\
&\quad=\int \mathrm{d}^2x\,\mu\big(\partial_tc+\nabla\cdot(c\mathbf{v})\big)
\\
&\quad=
\int\mathrm{d}^2x\,\nabla\mu\cdot\mathbf{q}_c-\int\mathrm{d}s\,\mu\,\mathbf{q}_c\cdot\mathbf{n}
\end{aligned}    
\end{equation}
The third term in~\eqref{eq:dissip1} vanishes on account of the constraint~\eqref{eq:incompressibility}. For the fourth term, standard shape-calculus identities (see e.g. \cite{Brummelen:2017sh}) lead to
\begin{equation}
\mathrm{d}_t\int{}\mathrm{d}s\,\gamma_s
=
-\int\mathrm{d}s\,\gamma_s\,\kappa\,\mathbf{v}\cdot\mathbf{n}
+
\gamma_s\,(\mathbf{t}_-+\mathbf{t}_+)\cdot\mathbf{v}(\mathbf{x}_{\text{cl}})
\end{equation}
It is to be noted that the surface integral in the first term extends across the
smooth parts of the interface on either side of the contact line. 

For the work functional~\eqref{eq:work}, it holds that
\begin{equation}
\label{eq:dtW}
\mathrm{d}_t\mathcal{W}[\boldsymbol{\psi}]=  
\gamma_{\textsc{lv}}\mathbf{t}_{\textsc{lv}}\cdot\partial_t\boldsymbol{\psi}(\mathbf{X}_{\text{cl}})
=
\gamma_{\textsc{lv}}\mathbf{t}_{\textsc{lv}}\cdot\mathbf{v}(\mathbf{x}_{\text{cl}})
\end{equation}
The final expression in~\eqref{eq:dtW} follows from transporting the penultimate expression from the reference configuration to the current configuration. 

Collecting the results in~\eqref{eq:dissip1}--\eqref{eq:dtW}, one obtains
\begin{equation}
\begin{aligned}
\mathrm{d}_t(\mathcal{E}-\mathcal{W})
&=
-\int\mathrm{d}^2x\,\boldsymbol{\sigma}_{\mathrm{v}}:\nabla\mathbf{v}
+\int\mathrm{d}^2x\,\nabla\mu\cdot\mathbf{q}_c
\\
&\phantom{=}
+\int\mathrm{d}s\,\mathbf{v}\cdot(\boldsymbol{\sigma}\cdot \mathbf{n}-\gamma_s\kappa\mathbf{n})
-\int\mathrm{d}s\,\mu\,\mathbf{q}_c\cdot\mathbf{n}
\\
&\phantom{=}
+\big(\gamma_s\,(\mathbf{t}_-+\mathbf{t}_+)-\gamma_{\textsc{lv}}\mathbf{t}_{\textsc{lv}}\big)
\cdot\mathbf{v}(\mathbf{x}_{\text{cl}})
\end{aligned}
\end{equation}
The dissipation relation~\eqref{eq:dedt} follows by introducing the constitutive relations~\eqref{eq:qc} and~\eqref{eq:viscoelastic}, the dynamic interface condition~\eqref{eq:dyncon}, the homogeneous Dirichlet condition for the chemical potential at the interface, and the Neumann law~\eqref{eq:Neumann}.

\bibliography{apssamp}

\end{document}